# MULTI-ENERGY PROTON EVENTS AND GEOMAGNETIC STORMS IN SOLAR CYCLES 23 AND 24

Rositsa Miteva[1], Susan W. Samwel[2], Svetoslav Zabunov[3]

[1]*Institute of Astronomy with National Astronomical Observatory – Bulgarian Academy of Sciences*
[2]*National Research Institute of Astronomy and Geophysics (NRIAG) – Egypt*
[3]*Space Research and Technology Institute – Bulgarian Academy of Sciences*
*e-mail: rmiteva@astro.bas.bg*

**Abstract:** *Solar energetic protons (SEPs) in different energy channels from 10 to above 100 MeV are analyzed and their relationship to solar and geomagnetic activity is investigated. We performed temporal association analysis between the SEPs, solar flares (SFs), coronal mass ejections (CMEs) and geomagnetic storms (GSs) that occurred during solar cycles 23 and 24. The energy dependencies between the SEPs and the strength of the space weather activity are evaluated and presented.*

# МУЛТИ-ЕНЕРГЕТИЧНИ ПРОТОННИ СЪБИТИЯ И ГЕОМАГНИТНИ БУРИ ПРЕЗ СЛЪНЧЕВИ ЦИКЛИ 23 И 24

Росица Митева[1], Сюзан В. Самуел[2], Светослав Забунов[3]

[1]*Институт по астрономия с национална астрономическа обсерватория – Българска академия на науките*
[2]*Национален изследователски институт по астрономия и геофизика – Египет*
[3]*Институт за космически изследвания и технологии – Българска академия на науките*
*e-mail: rmiteva@astro.bas.bg*

*Ключови думи: Слънчеви енергетични протони, геомагнитни бури, слънчеви избухвания, коронални изхвърляния на маса*

*Резюме: Слънчеви енергетични протони в различни канали от 10 до над 100 MeV са анализирани и техните взаимовръзки със слънчевата и геомагнитната активност са изследвани. Извършен е времеви анализ между протонните събития, слънчевите избухвания, короналните изхвърляния на маса и геомагнитните бури за времето на слънчеви цикли 23 и 24. Зависимостите по енергия между частиците и силата на проявите на космическо време са намерени и представени.*

### Introduction

Geomagnetic storms (GSs) are disturbances in the terrestrial magnetosphere due to the energy input from the incoming solar wind or magnetized plasma structures with southward directed magnetic field (Dungey, 1961). Different current systems are subsequently being generated by flows of energetic particles that are employed to describe the different phases of the GS (e.g. Akasofu, 2018): the current enhancement on the magnetospheric boundary surface causes the storm sudden commencement (small increase) and the storm initial (decrease) phase; the magnetospheric ring current around the Earth is responsible for the storm main (decrease) phase; and the intense ionospheric currents (so-called auroral electrojet) lead to polar disturbances (auroras).

The strength of the storm is traditionally defined by the disturbance storm time (Dst) index, measuring the decrease of the Earth magnetic field in nT on ground level by a set of equatorial magnetometers. Other indices also exist and are occasionally explored (see e.g. Vennerstrom et al. 2016). In general it is considered, that stronger storms, also occurring in the maximum of the solar cycle (SC), are more often due to magnetic clouds (MC), whereas weaker storms, in the SC minimum, are due to co-rotating interacting regions (see Tsurutani et al. 2020 and the references therein). High-speed solar wind streams are used to explain the (usually weaker) 27-day-recurrent GSs.

The origin of the GS needs to be sought at the Sun, either in terms of eruptive phenomena or changes in the speed and density of the solar wind. The solar activity phenomena are traditionally regarded in terms of solar flares (SFs) and coronal mass ejections (CMEs). Both are the impulsive evidence of the solar magnetism, originally evidenced since ancient times as sunspots, located in the so-called active regions (ARs). A SF is the energy release of the stored magnetic energy via process of magnetic reconnection (e.g., Benz, 2016) and simplistically could be regarded as the 'flash' or a localized intense brightening observed on solar images, often producing saturations. This is why the SF is represented by the peak value in soft X-rays, denoted as flare class and measured in W/m$^2$. Apart from the light emitted over the entire electromagnetic spectrum (during large flares), a SF is accompanied by mass motion, acceleration of particles, rearrangement of magnetic field lines. Closely related to the SFs (especially large ones) are the CMEs. These are large volumes of the solar corona, plasma and embedded magnetic field, expelled in the interplanetary (IP) space (e.g. Webb and Howard 2012) often leading to substantial restructuring in the solar corona and disturbances in the IP medium. In coronagraph images, CMEs look like 'bubbles' of various shapes and sizes. The projected, on-sky measured linear speed (km/s), is the most representative parameter of the CME, together with their angular width (degrees). The IP counterparts of the CMEs (so-called ICMEs, Kilpua et al. 2017) and a subset of them – MCs, both observed by in situ instruments when arriving at Earth, are frequently accompanied by IP shock waves and energetic storm particles and are canonically regarded as the drivers of (intense) GSs.

Solar energetic particles (SEPs) are protons, ions or/and electrons accelerated close to the Sun, at their so-called origin – SFs or/and CMEs, that escaped from the solar corona along open magnetic field lines, and are transported along the IP field lines to Earth where they are observed 'in situ' by dedicated satellite instruments (e.g., Desai and Giacalone 2016). There is a rich literature on the topic of SEPs, their solar origin and their associated phenomena as well as their space weather relevance (e.g., Schwenn, 2006, Pulkinnen, 2007).

The aim of the present study is to investigate the relationship between GSs and their accompanied space weather phenomena, CMEs, SFs and SEPs. Due to their earlier arrival at Earth compared to the GS, the SEP occurrence and energy is tested for being a reliable GS precursor. The geo-effectiveness of the SEPs is previously investigated in SC23 by e.g. Le et al. (2016), whereas here we consider also SC24, start with a list of all GSs with Dst ≤ −100 nT and focus on the multi-energy observations of solar protons.

**Data analysis**

The GSs in the period of interest (1996–2019) are selected from the Kyoto Geomagnetic data service (http://wdc.kugi.kyoto-u.ac.jp/wdc/Sec3.html), as provided for following three periods:
1957-2014: http://wdc.kugi.kyoto-u.ac.jp/dst_final/
2015-2016: http://wdc.kugi.kyoto-u.ac.jp/dst_provisional/
2017-now: http://wdc.kugi.kyoto-u.ac.jp/dst_realtime/

After a visual cross-check of the quick-look plots and textual reports in the above databases, in total of 107 GSs with Dst ≤ −100 nT are found. The reported time there is rounded to the nearest hour.

For the analysis here, each GS is temporally associated with a CME, SF and SEP, where possible. The following set of criteria is applied:
- The parent CME is initially sought in a three-day time window prior the GS onset time. In case of multiple CMEs, we estimate the individual transit times by means of the reported linear CME speeds (assumed constant during propagation in the IP space) and adopt the one whose arrival time at Earth is closest to the GS onset time (e.g., Miteva, 2020a). If no suitable candidate is found, the search window is extended. The identified CME (and SF) are compared with the associations given by Zhang et al. (2007) in SC23 and by Singh et al. (2017) during the ascending phase of SC24. Our CME association is also compared with the reported LASCO CME in the ICME online catalog by I. Richardson and H. Cane: http://www.srl.caltech.edu/ACE/ASC/DATA/level3/icmetable2.html
- The so-identified CME is then associated with a SF that is closest in time and originating from the same solar quadrant or AR. The following well-known SF database is used: ftp://ftp.ngdc.noaa.gov/STP/space-weather/solar-data/solar-features/solar-flares/x-rays/goes/.
- Finally, the CME–SF pair is related to an in situ proton event as observed by the SOHO/ERNE instrument, under the requirement the solar activity to be the strongest and closest in time prior the SEP event (according to the association criteria adopted by Miteva et al. 2018, Miteva, 2019).
- For the SEP events we used the preliminary results for the proton amplitude as observed in the 10 channels of the high energy detector aboard the SOHO/ERNE instrument (Torsti et al. 1995), see Miteva (2019), Miteva et al. (2020b) (https://catalogs.astro.bas.bg/).

**Results**

After the performed association analyses we obtain that the 107 GSs are accompanied with 73 CMEs, 64 SFs and 35 different SEP events in the energy range 14–17 MeV, decreasing to 14 protons in the highest energy channel, 101–131 MeV.

Despite the adopted quantitative criteria for association, there are number of GSs with uncertain solar origin (34/107 with no CME, also due to data gaps, and 43/107 with no SF association). The distribution of the Dst index of these 107 GSs is shown in Fig. 1 (on the left) and the histogram for the subset of 73 GSs with identified CMEs are shown on the right. The mean (median) values of the Dst index are 155 (130) in the former case and ~170 (~145) in the latter. It is evident by comparing the plots in Fig. 1 that the solar origin association is easier to perform for the stronger GSs, as the bin of the weakest GS is shortened in half.

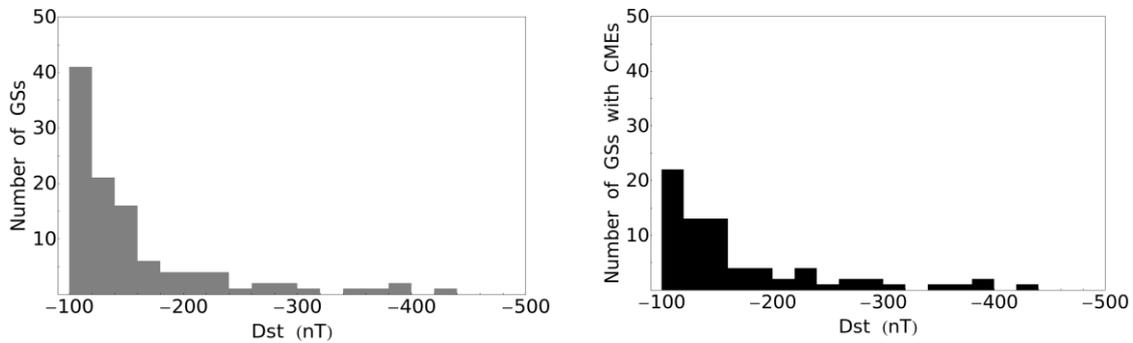

Fig. 1. Number of all GSs in the period 1996–2019 with Dst index ≤–100 nT (left plot) and a subset of the GSs with identified parent CMEs (right plot). The bin width is 20 nT.

In this study we start with a list of GSs and explore only this direction of association. Firstly, the relationship with the solar origin is presented in Fig. 2. Double logarithmic scatter plots between the Dst index of the GS and the SF class/CME speed, respectively, are used there. The distribution with the SF is random, whereas the one with CME speed shows some structure. Apart from the vertical cutoff at –100 nT (at both plots) pre-selected here for the GSs, we notice a nearly horizontal limit at the maximum 3000 km/s for the CME speed. For weak GSs, the spread in the parent CME speeds range from 100s to close to 3000 km/s, whereas for the large GSs only faster than 1000 km/s CMEs are obtained. This dependency is producing the roughly triangular shape of the distribution.

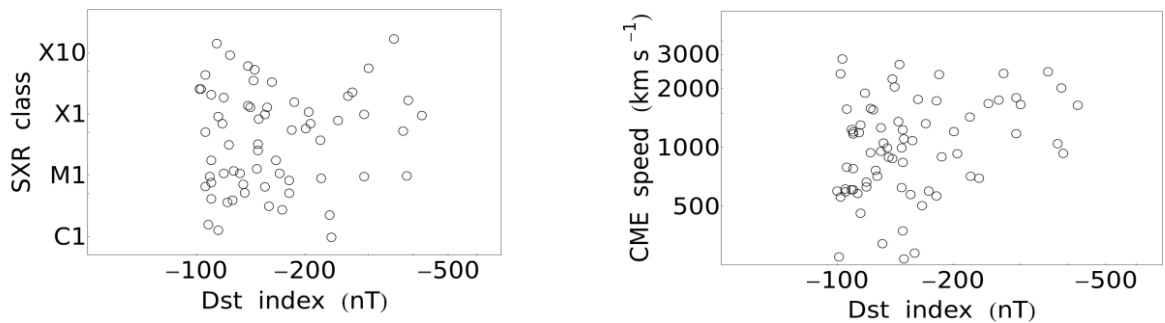

Fig. 2. Scatter plots between the GS Dst index and the SF class (on the left) and CME speed (on the right)

The scatter plots (in log-log form) between the GSs and the SEPs are presented in Fig. 3 for all 10 energy channels. The majority of the SEP events accompanied with GSs have larger peak flux than the average for the population (e.g., Miteva et al. 2020b). At the present study, only the observed proton fluxes are used, without performing a flux correction due to the instrument saturation effects.

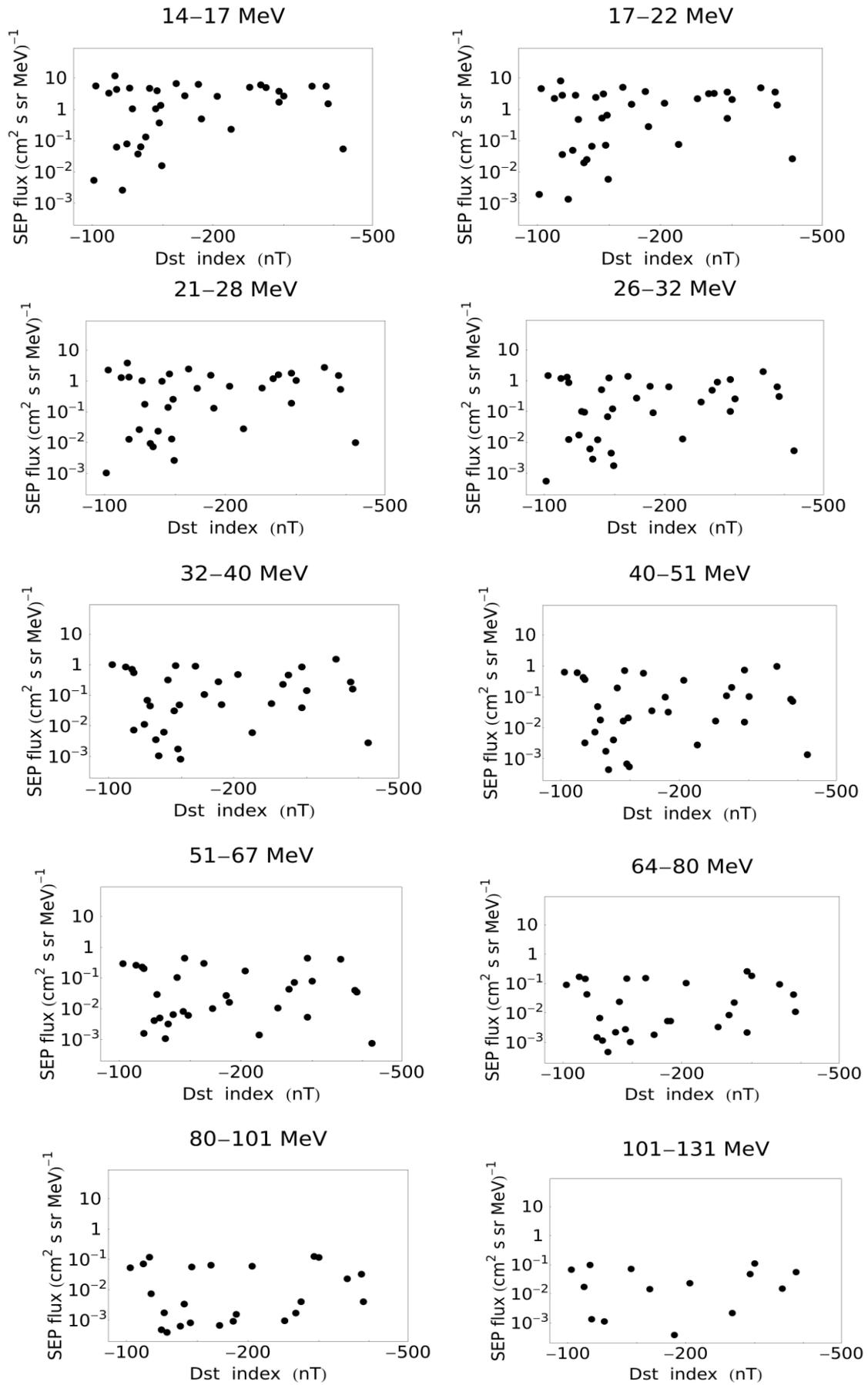

Fig. 3. Scatter log-log plots between the Dst index and the SEP peak flux for the different proton energy channels

The number of the associated pairs reduces slightly with the energy increase, as the number of SEPs naturally declines with increasing of their energy. For the first several energy channels (excluding the outlier for the largest GS in the sample accompanied with a moderate SEP flux), the distribution shape is similar to the one for the CMEs, whereas for the remaining 6, the distribution is more arbitrary.

**Summary**

In the present work, the geo-effectiveness of SEP events is investigated in terms of their relationship with GSs in SC23 and 24. Based on the analysis of GSs in this period, we found that GSs show the same relationship with low energy protons (in the 14–32 MeV range) as with their parent CMEs, whereas for the high energy protons there is no clear relationship to GSs strength, thus the geomagnetic prediction ability of the 32–131 MeV protons is limited. In contrast, the lower energy protons could well be used as a proxy for their parent CMEs while relating them to GSs. For completeness, a future analysis will explore the reverse direction of association (starting with a list of SEPs), in addition to employing the corrected SOHO/ERNE fluxes (Miteva et al. 2020b).

**Acknowledgements**

The analysis is partly supported by the Bulgarian National Science Fund under contracts KP-06-H28/4 (08-Dec-2018) and KP-06-India/14 (19-Dec-2019) and SCOSTEP/PRESTO grand for 2020 'On the relationship between major space weather phenomena in solar cycles 23 and 24'.